\documentclass[useAMS,usenatbib]{mn2e}
\usepackage{graphicx}
\begin{document}
\title[The Birthrate of Magnetars]{The Birthrate of Magnetars}

\author[R. Gill and J. Heyl]{Ramandeep Gill\thanks{Email: rsgill@phas.ubc.ca} and Jeremy Heyl$^{1}$\thanks{Email: heyl@phas.ubc.ca; Canada Research Chair}\\
$^{1}$Department of Physics and Astronomy, University of British Columbia \\
6224 Agricultural Road, Vancouver, British Columbia, Canada, V6T 1Z1}

\date{\today}

\pagerange{\pageref{firstpage}--\pageref{lastpage}} \pubyear{2007}

\maketitle

\label{firstpage}

\begin{abstract}
  Magnetars, neutron stars with ultra strong magnetic fields ($B\sim
  10^{14} - 10^{15}$G), manifest their exotic nature in the form of
  soft gamma-ray repeaters and anomalous X-ray pulsars. This study
  estimates the birthrate of magnetars to be $\sim$ 0.22 per century
  with a galactic population comprising of $\sim$17 objects. A
  population synthesis was carried out based on the five anomalous
  X-ray pulsars detected in the ROSAT All-Sky Survey by comparing
  their number to that of massive OB stars in a well defined
  volume. Additionally, the group of seven X-ray dim isolated neutron
  stars detected in the same survey were found to have a birthrate of
  $\sim$ 2 per century with a galactic population of $\sim$ 22,000
  objects.
\end{abstract}


\section{Introduction}

The end products of massive stars with initial masses in the range
$8M_{\odot} < M_{initial} < 25M_{\odot}$, neutron stars (NSs)
constitute $\sim 0.0001\%$ of the total stellar population of the
Milky Way in the form of active pulsars \citep{lynesmith2006}. It is
now understood that not all NSs turn out to exhibit radio
pulsations. This missing link manifests itself in a highly energetic
class of NSs called magnetars \citep{zhang2002}. \citet{dt1992}
proposed the existence of magnetars with ultra strong magnetic fields
100 times stronger than that observed for normal radio pulsars ($B\sim
10^{12}$ G; see \citealp{ziolkowski2002}). Magnetar formation, they
argued, is caused by a vigorous helical dynamo action during the
initial stages of birth for NSs with initial periods $\sim 1$
ms. Furthermore, the magnetar model predicts the occurrence of
gamma-ray or X-ray bursts from these powerful sources due to
starquakes caused by strong magnetic field activity \citep{dt1992,
  dt1996}.

Such bursting behavior has been observed for both soft gamma-ray
repeaters (SGRs) and anomalous X-ray pulsars (AXPs; see
\citealp{kaspi2003,gavriil2002}). On March 5, 1979, a high intensity
burst (similar to a short gamma-ray burst) and subsequent smaller
bursts from the same patch of sky recorded by several orbiting
spacecrafts unveiled the existence of soft gamma-ray repeaters in our
galaxy \citep {mazets1979}. Interestingly, the burst from this source
only lasted $\sim 0.1$ s, but was followed by a decaying less
energetic periodic pulse of period $\sim 8$ s
\citep{terrell1980}. Later this source was recognized as SGR 0525-66
in the supernova remnant (SNR) N49, outside our galaxy in the Large
Magellanic Cloud. Following this discovery, the rest of the known SGRs
(SGR 1806-20, SGR 1900+14, SGR 1627-41) were discovered in our galaxy
over a period of 25 years. Subsequent observations of SGR 1806-20
confirmed that SGRs are indeed magnetars characterized by high
quiescent X-ray luminosities and strong magnetic fields
\citep{kouveliotou1998}.

The AXPs are another class of very energetic and young NSs now
believed to be associated to SGRs, and are also considered magnetars
(see \citealp{mereghetti2002}). As of yet only 7 confirmed AXPs have
been discovered, namely 1E 2259+586, 4U 0142+61, 1E 1048.1-5937, 1RXS
J170849.0-400910, 1E 1841-045, CXOU J010043.1-721134, and XTE
J1810-197\footnote{http://www.physics.mcgill.ca/$\sim$pulsar/magnetar/main.html}. Both
SGRs (in quiescence) and AXPs share many similarities and are
generally characterized by steady X-ray luminosities ($L_x\sim 10^{32}
- 10^{35}\mbox{ergs s}^{-1}$), long spin periods ($P\sim 5 - 12$ s),
secular spin down rates with large period derivatives ($\dot{P}\sim
10^{-11}\mbox{ s s}^{-1}$), strong magnetic fields ($B\sim 10^{14} -
10^{15}$ G), and an apparent lack of a binary companion
\citep{wt2006}. Nevertheless, some spectral differences exist between
the two. Although pulsed radio emission has been detected from only
one of the magnetars (AXP J1810-197; \citealp{camilo2006}), the rest
are found to be radio quiet.

In addition, two new groups of NSs, namely the X-ray dim isolated NSs
(XDINSs) and the rotating radio transients (RRATs), have recently been
discovered. The XDINSs are a group of seven radio-quiet NSs (often
referred to as the `magnificent seven') associated to a nearby OB star
complex comprising the Gould belt \citep{popov2003}. These objects
have spin periods between 3 - 11 s, with relatively low X-ray
luminosities ($L\sim 10^{30}-10^{31}\mbox{ergs s}^{-1}$) and magnetic
fields ($B\sim 10^{13}$ G) compared to the SGRs/AXPs (see
\citealp{haberl2004,haberl2005,treves2000}). Here the reader is
cautioned not to generalize the magnetic field estimate to the whole
XDINS family as it is solely derived from the spin period and period
derivative measurements of just two of its members (RX J0720.4-3125,
RBS 1223; \citealp{kaplan2005a,kaplan2005b}). For the remaining XDINSs
no period derivatives have been obtained so far. As the name suggests,
RRATs are transient radio sources that exhibit sporadic radio bursts
of duration between 2 - 30 ms \citep{mclaughlin2006}. These radio
bursts have a maximum flux density of $\sim 0.1-0.4$ Jy with burst
time intervals between 4 min - 3 hr. What makes RRATs worthy of
consideration is that their spin periods ($\sim 0.4-7$ s) and magnetic
fields ($B\sim 10^{13}$ G) are just in the right range to allow
suspicion of some association between these and the XDINSs
\citep{popov2006}. So far 11 such sources have been discovered.

To make the scene even more congested, yet another class of NSs begs
its introduction. Characterized by long periods ($\sim$ 0.2 - 7 s),
high magnetic fields ($\sim 10^{13}$ G), and spindown ages $\sim 10^{3
  - 4}$ yrs, high B-field radio pulsars (HBRPs) are strong candidates,
among the above said objects, in having any connection with the
magnetar population \citep{gonzalez2006,vranesevic2007}. Many members
of this class have been detected in the X-rays while others have
displayed exceedingly dim X-ray emission.

Due to the paucity of sources, no population synthesis of magnetars
has been performed. As a result, it is still unclear how magnetars fit
into the grand scheme. Basically, the questions that we are asking in
this study are:
\begin{enumerate}
\item
What is the birthrate of magnetars?
\item
How these objects are associated to other classes of NSs?
\end{enumerate}
\citet{kouveliotou1998} argued that roughly 10\% of all NSs become
magnetars and that their birthrate is high $\beta_{Mag}\sim 0.1$ per
century. Estimates of the same magnitude have been proposed based on
the spin down histories of AXPs or derived solely from the age of the
SNR to which few magnetars are associated \citep{paradijs1995}. In
other instances it has also been noted that spindown ages of magnetars
may not strictly correspond to the age of the SNRs due to decaying
fields \citep{colpi2000}. This renders any birthrate calculations to
be unsatisfactory since these assume constant magnetic
fields. Moreover, \citet{wt2006} conjecture that birthrate estimates
are dependent on the efficiency of finding these magnetars which is
limited by the sensitivity of the detector. Therefore, it is necessary
to conduct a population synthesis in order to obtain a reasonable
(possibly a lower limit because of transients) estimate of the
birthrate of magnetars. Here the authors deem necessary to stress the
point that this study only looks at the persistent AXPs and no
transient sources are included in the source sample, thus, only a
lower limit to the magnetar birthrate can be obtained.

The rest of the paper is organized as follows: Section 2 outlines the
method employed for this study with a brief discussion of the ROSAT
All-Sky Survey. The results and the corresponding discussion are
presented in Section 3. Finally, we end with a brief note of
recommendation in Section 4 and the conclusion of this study in
Section 5.
\section{Method}

To provide an overview of the analysis, the whole scheme of
calculation has been divided into three main tasks. First, the
detector count rate at a fiducial distance of 1 kpc is calculated as a
function of the column density $N_H$, for each object in the
sample. The sample comprises of the five AXPs and the seven XDINSs,
all detected in the ROSAT All-Sky Survey (RASS) with a high S/N
ratio. The SGRs were not detected in the statistically well-defined
RASS so they are not included in the sample.  Rather we have simply
assumed that the existing cadre of instruments could have detected any
active SGR in the Galaxy. Second, the maximum distance to which a
source registered a count rate of (a) $\geq$ 0.05 cts $\mbox{s}^{-1}$
and (b) $\geq I_{min}$ (the ROSAT count rate with the lowest
acceptable S/N ratio) in the RASS, is computed. The maximum distance
is expected to vary with position on the sky due to the difference in
the integrated column densities; peering into the galactic disk is
much harder in comparison to the galactic poles. Third, this maximum
distance is then transformed into a limiting volume within which the
number of massive OB stars, that are the most likely progenitors of
neutron stars, is sought. An exercise of this sort requires the use of
a smoothly varying luminosity function as well as the number density
of stars in a given volume. Comparison of the number of massive OB
stars found in the surveyed volume to that of their total population
yields the scaling factor. The scaled populations of AXPs and XDINSs
complemented by the knowledge of their spindown ages, finally, provide
their birthrates in the Galaxy.
\subsubsection*{The ROSAT All-Sky Survey (RASS)}

ROSAT, short for ROentgen SATellite, was an X-ray space observatory
launched on June 1, 1990. Equipped with a position-sensitive
proportional counter (PSPC), a $2^{\circ}$ field of view, ROSAT
produced an unprecedented all sky survey in a period of 6 months. It
surveyed the whole sky in the $\sim 0.12 - 2.4$ keV ($\sim 100\mbox{\AA} -
6\mbox{\AA}$) energy range at a typical limiting count rate of 0.015 cts
$\mbox{s}^{-1}$ \citep{hunsch1999}. ROSAT had an orbital period of
$\sim$ 96 min with 40 min nighttime per orbit and it scanned the whole
sky in great circles perpendicular to the line connecting the Earth
and the Sun. Consequently, regions on the sky falling at the ecliptic
registered much less exposure time (typically $\sim$ 400 s) than that
at the ecliptic poles ($\sim$ 40,000 s). Besides this artifact,
introduced by the geometry of the survey, each RASS field suffered
from inconsistent exposure time variations. This is illustrated in
Fig~1 where the pixels of each field registered a wide range of
exposure times between 200 s and a little over 650 s.
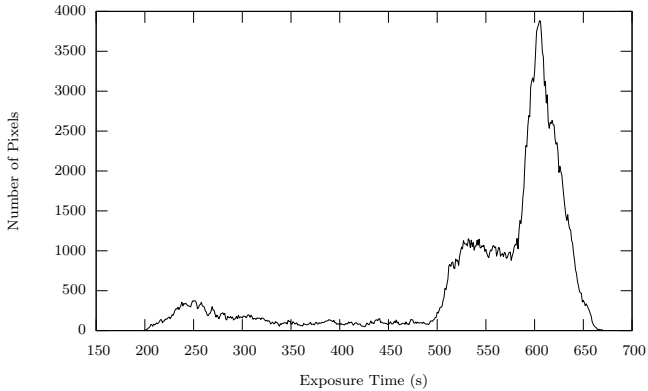
\begin{figure}
\begin{center}
\scalebox{0.7}{
%GNUPLOT: LaTeX picture with Postscript
\begin{picture}(0,0)%
\includegraphics{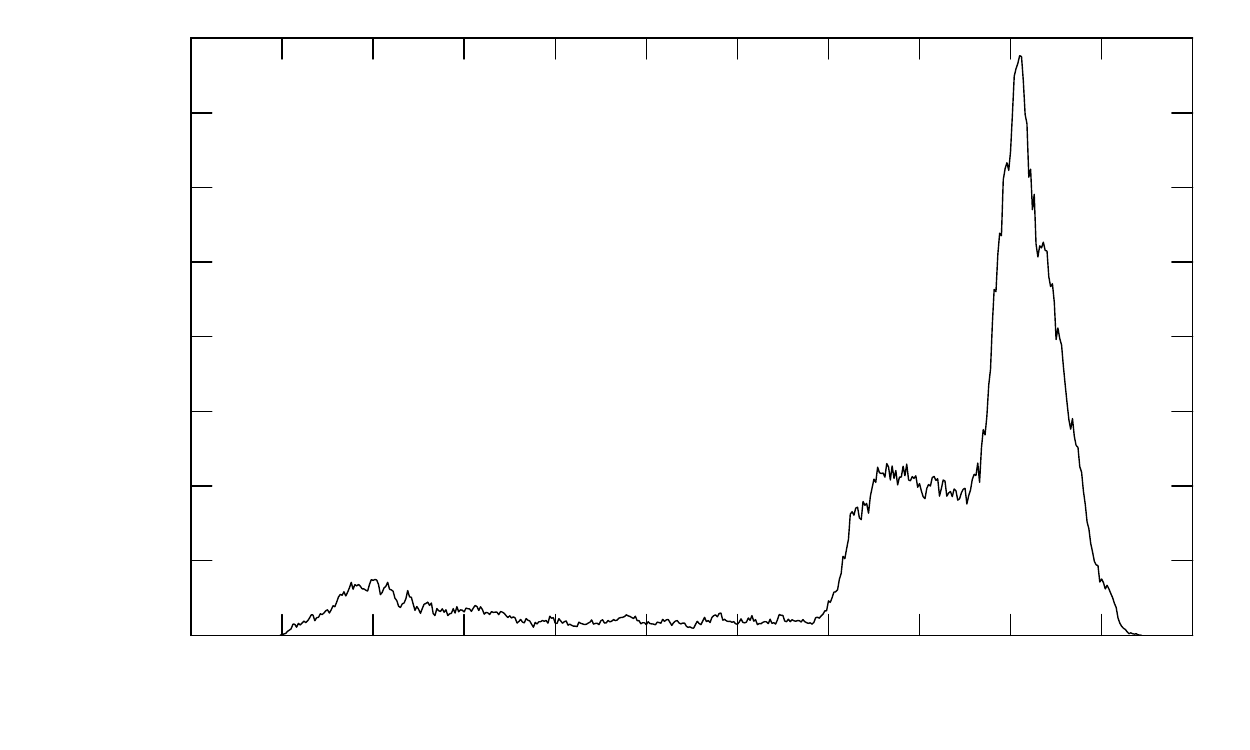}%
\end{picture}%
\begingroup
\setlength{\unitlength}{0.0200bp}%
\begin{picture}(18000,10800)(0,0)%
\put(2475,1650){\makebox(0,0)[r]{\strut{} 0}}%
\put(2475,2725){\makebox(0,0)[r]{\strut{} 500}}%
\put(2475,3800){\makebox(0,0)[r]{\strut{} 1000}}%
\put(2475,4875){\makebox(0,0)[r]{\strut{} 1500}}%
\put(2475,5950){\makebox(0,0)[r]{\strut{} 2000}}%
\put(2475,7025){\makebox(0,0)[r]{\strut{} 2500}}%
\put(2475,8100){\makebox(0,0)[r]{\strut{} 3000}}%
\put(2475,9175){\makebox(0,0)[r]{\strut{} 3500}}%
\put(2475,10250){\makebox(0,0)[r]{\strut{} 4000}}%
\put(2750,1100){\makebox(0,0){\strut{} 150}}%
\put(4061,1100){\makebox(0,0){\strut{} 200}}%
\put(5373,1100){\makebox(0,0){\strut{} 250}}%
\put(6684,1100){\makebox(0,0){\strut{} 300}}%
\put(7995,1100){\makebox(0,0){\strut{} 350}}%
\put(9307,1100){\makebox(0,0){\strut{} 400}}%
\put(10618,1100){\makebox(0,0){\strut{} 450}}%
\put(11930,1100){\makebox(0,0){\strut{} 500}}%
\put(13241,1100){\makebox(0,0){\strut{} 550}}%
\put(14552,1100){\makebox(0,0){\strut{} 600}}%
\put(15864,1100){\makebox(0,0){\strut{} 650}}%
\put(17175,1100){\makebox(0,0){\strut{} 700}}%
\put(550,5950){\rotatebox{90}{\makebox(0,0){\strut{}Number of Pixels}}}%
\put(9962,275){\makebox(0,0){\strut{}Exposure Time (s)}}%
\end{picture}%
\endgroup
}
\caption{Variable exposure time of an individual ROSAT field. The
  majority of the pixels registered exposure times between 500 and 650
  s. This produced variations in the sensitivity limit of the detector
  for each point on the sky. Thus, the limiting count rate is a
  function of the exposure time.}
\end{center}
\end{figure} 
\subsection{Count Rate Vs HI Column Density}

To simulate ROSAT observations of the AXPs in the 0.1 - 2.4 keV energy
range, a two component model (Powerlaw + Blackbody) modulated by
interstellar absorption is used. Since the XDINSs only exhibit thermal
spectra with $kT\la 100$ eV \citep{popov2006}, only the blackbody
component is implemented for these objects.
\begin{equation}
A(E) = \left[\eta_1\left(\frac{E}{1\mbox{ keV}}\right)^{-\alpha} + \eta_2\frac{8.0525E^2dE}{k^4T^4\{\exp\left(\frac{E}{kT}\right)-1\}}\right]
\end{equation}
\[
\hspace{15em}\times\exp[-N_H\sigma(E)]
\]
In the above, $\eta_1$ is the powerlaw normalization with units of
photons$\mbox{ keV}^{-1}\mbox{cm}^{-2}\mbox{s}^{-1}$, $\eta_2$ is the
blackbody normalization with units of luminosity per square distance
($L_{39}/D_{10}^2$; with $L_{39}$ the luminosity in units of
$10^{39}\mbox{ ergs s}^{-1}$ and $D_{10}$ the distance in units of 10
kpc). With all the spectral parameters known the model has to be
properly normalized for each object. Model normalizations for AXPs 4U
0142+61, 1E 1048.1-5937, and 1RXS J170849.0-400910 are obtained from
\citet{perna2001}. Normalizations are not available for AXPs 1E
2259+586 and 1E 1841-045, so we use the percentage fluxes of both
model components contributing to the total unabsorbed flux of the
source \citep{patel2001,morii2003}. We then normalize the simulated
absorbed fluxes to that recieved at the telescope to obtain the right
count rate at the source specific $N_H$. Since XSPEC only provides
flux estimates, an ftools subprogram PIMMS is used to convert flux
into ROSAT count rate. Figure 2 displays the log of the simulated
count rate as a function of $N_H$ at a fiducial distance of 1 kpc for
AXP 4U 0142+615 (actual distance of source is 3 kpc). In the figure,
the exponential attenuation of the count rate with increasing column
density is very much apparent. A sudden change in the slope at low
column densities is due to the absorption of softer photons.
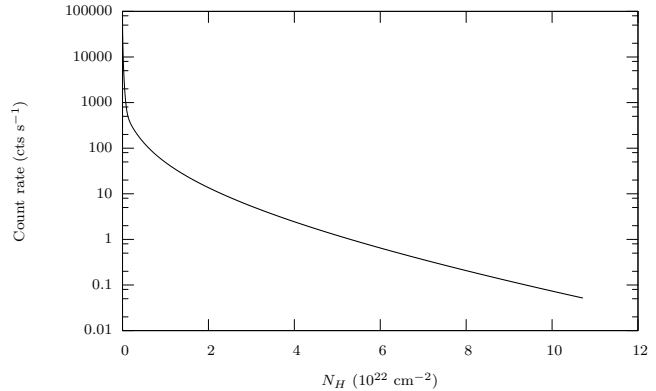
\begin{figure}
\begin{center}
\scalebox{0.7}{
%GNUPLOT: LaTeX picture with Postscript
\begin{picture}(0,0)%
\includegraphics{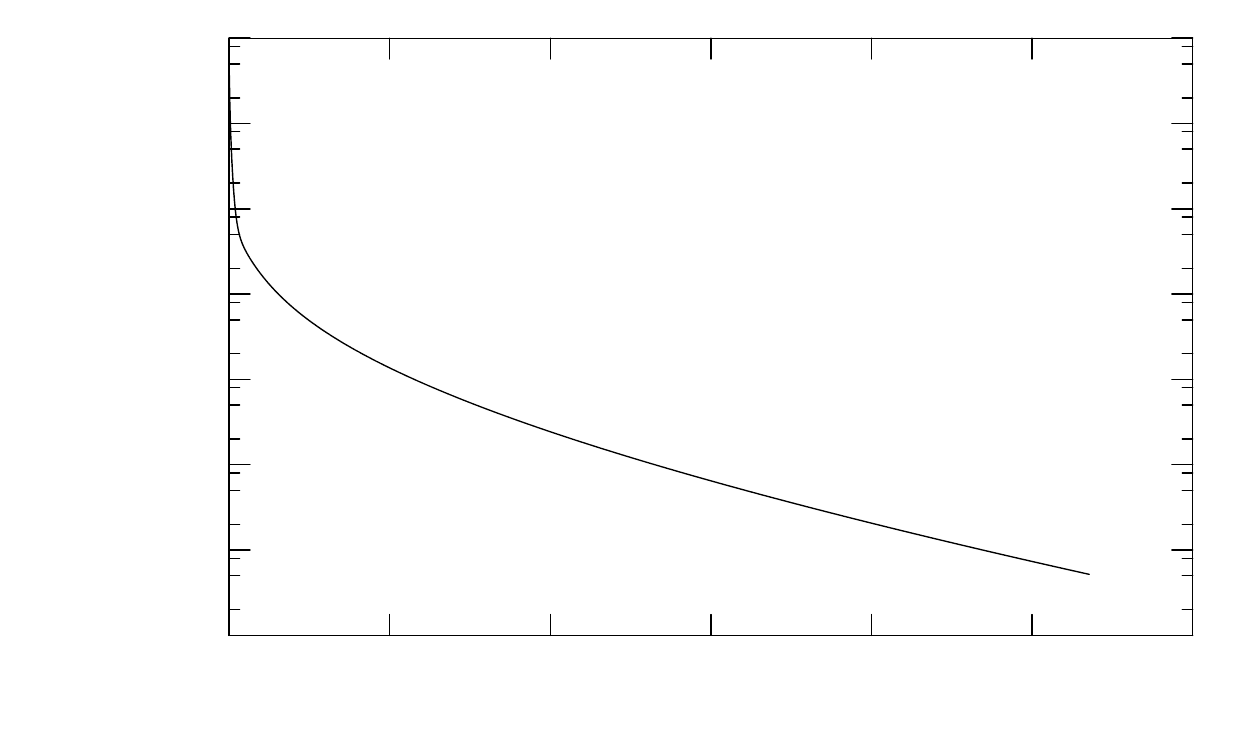}%
\end{picture}%
\begingroup
\setlength{\unitlength}{0.0200bp}%
\begin{picture}(18000,10800)(0,0)%
\put(3025,1650){\makebox(0,0)[r]{\strut{} 0.01}}%
\put(3025,2879){\makebox(0,0)[r]{\strut{} 0.1}}%
\put(3025,4107){\makebox(0,0)[r]{\strut{} 1}}%
\put(3025,5336){\makebox(0,0)[r]{\strut{} 10}}%
\put(3025,6564){\makebox(0,0)[r]{\strut{} 100}}%
\put(3025,7793){\makebox(0,0)[r]{\strut{} 1000}}%
\put(3025,9021){\makebox(0,0)[r]{\strut{} 10000}}%
\put(3025,10250){\makebox(0,0)[r]{\strut{} 100000}}%
\put(3300,1100){\makebox(0,0){\strut{} 0}}%
\put(5613,1100){\makebox(0,0){\strut{} 2}}%
\put(7925,1100){\makebox(0,0){\strut{} 4}}%
\put(10238,1100){\makebox(0,0){\strut{} 6}}%
\put(12550,1100){\makebox(0,0){\strut{} 8}}%
\put(14863,1100){\makebox(0,0){\strut{} 10}}%
\put(17175,1100){\makebox(0,0){\strut{} 12}}%
\put(550,5950){\rotatebox{90}{\makebox(0,0){\strut{}Count rate ($\mbox{cts s}^{-1}$)}}}%
\put(10237,275){\makebox(0,0){\strut{}$N_H$ ($10^{22}\mbox{ cm}^{-2}$)}}%
\end{picture}%
\endgroup}
\caption{Log of count rate as a function of $N_H$ at a fiducial distance of 1 kpc for AXP 4U 0142+61. The sharp drop off is due to the exponential attenuation of the source flux with increasing $N_H$ along the line of sight. Most AXPs, in this study, have measured column densities of $\sim10^{22}\mbox{ cm}^{-2}$. Conversely, XDINSs have much smaller column densities due to their nearness.}
\end{center}
\end{figure}
\subsection{Maximum Probed Distance}
The sensitivity limit of a telescope and the duration for which it was exposed to the sky determines how dim, or how far, an object could be detected. What is sought here is the distance to which a source with known luminosity could be detected by ROSAT before its flux drops below the detector sensitivity threshold. In this study, two detection criteria have been selected. First, a limiting count rate of 0.05 cts $\mbox{s}^{-1}$ (criterion A) is chosen, for it conforms to the standards of the RASS bright source catalog \citep{voges1999}. All five AXPs appear in the RASS-BSC and, thus, are well above the standardized detection threshold of a limiting count rate of $0.05\mbox{ cts s}^{-1}$, a detection likelihood\footnote{Detection likelihood is defined as $-\ln(1-p)$ where $p$ is the probability of source existence.} of at least 15, and an accumulation of at least 15 source photons. Second, an exposure-time dependent minimum count rate (criterion B) is chosen since exposure time variation across each RASS field alters the detector threshold in the following way
\begin{equation}
I_{min} = \frac{1}{\sqrt{t}}\left[\frac{N_{\sigma}^2 B_{cell}}{f^2} + \frac{N_{min}^2}{t}\right]^{\frac{1}{2}}
\end{equation}
with
\begin{equation}
B_{cell} = \left(\frac{A_{cell}}{A_{FOV}}\right)B_{FOV}
\end{equation}
In the above, $N_{\sigma}=5$ is the signal to noise ratio, $B_{cell}$ is the background count rate in the source detection cell, $A_{cell}$ is the area of the detection cell which, in this study, was assumed to be the same as the source extraction area (typically $\sim 300 \mbox{ arcsec}^2$), $A_{FOV}$ is the area of the field of view of the telescope, $f=0.9$ is the fraction of source counts in the detection cell, $N_{min}=15$ is the minimum number of photons required to make a detection, and $t$ is the integrated exposure time. The value for $B_{FOV}$ is obtained from the all sky background count rate map in the 0.4 - 2.4 keV energy range.
\subsubsection*{Neutral Hydrogen Model}
The amount of HI integrated along the line of sight limits the distance to which a source could be observed. Generally, the HI distribution takes the form of an exponentially decaying disk, both radially and vertically, away from the galactic center. In principle, it should be modified with a warp and an additive spiral arm component \citep{bm1998}. For the purpose of this study, a simple exponentially varying distribution, of the form given below, is adopted from \citet{fosterroutledge2003}.
\begin{equation}
n(r,l,b) = n_{\circ}\exp\left[-\frac{R(r,l,b) - R_{\circ}}{h_R}-\frac{Z_{\odot} + r\sin |b|}{h_z}\right]
\end{equation}
with
\begin{eqnarray}
R(r,l,b) &=& R_{\circ}^2 + \left(r\cos b\right)^2 - 2R_{\circ}r\cos b\cos l\\
h_z &=& h_{\circ} + h'[R(r,l,b) - R_{\circ}]
\end{eqnarray}
where $h_R = 0.35R_{\circ}$, $h_{\circ} = 200\mbox{ pc }$, $h' = 30\mbox{ pc kpc}^{-1}$, and $R_{\circ} = 8.0\mbox{ kpc}$. Notice in (6), the vertical scale height is a linear function of the galactocentric radius $R$. \citet{fosterroutledge2003} argue that for lines of sight probing the very midplane of the galaxy, it is proper to adopt such a dependence as observations confirm a thickening of HI layer beyond the solar circle. In order to have a manageable resolution and to keep simulation times shorter, the whole sky is divided into 1440$\times $720 points over Right Ascension (RA) and Declination (DEC) spanning $360^{\circ}$ and $180^{\circ}$, respectively, with $0.25^{\circ}$ intervals. Assuming a reasonable value for $n_{\circ}=0.5\mbox{ cm}^{-3}$, the local number density of HI, and integrating (4) to a heliocentric distance $r$, for a given galactic position ($l,b$), yields the HI column density.
\begin{equation}
N_{H}(r,l,b) = \int_0^r n(r',l,b)dr'
\end{equation}
The $N_H$ inferred from (7) corresponds to a ROSAT count rate from the analysis in the previous section, illustrated by Fig 2. Until the count rate fails to satisfy the detection criteria, the integral in (7) is iterated with an increasing heliocentric distance. Finally, a maximum distance to which a given source, in a given direction, could be detected by ROSAT is obtained. This exercise is repeated for all twelve sources in the sample as well as for both count rate criteria.
\subsection{Starcounts}
In a magnitude limited sample, containing stars brighter than a given absolute magnitude $M$, star counts are usually calculated by assuming some stellar luminosity function $\Phi_M(M)$, and a model of stellar distribution $n_{\star}(r, l, b)$ in the galaxy. Consider a sample of stars with absolute magnitudes between $M$ and $M+dM$, lying in a specific direction ($l,b$), in a given solid angle $\Delta\Omega$, within some distance $d$. Then their total number is simply given by the following expression \citep{bm1998}.
\begin{equation}
N(\Omega, d, l, b) = \int_{M_1}^{M_2}\Phi_M(M)dM \int_0^d n_{\star}(r, l, b) r^2 \Delta\Omega dr
\end{equation}
We adopt $M_1 = -4.9$ for the brightest stars and $M_2=-1.6$ for the dimmest stars that yield neutron stars. This particular choice only affects the normalization of the star counts but not the derived birth rates which depend on the assumed value of $n_{\star}(r,l,b)$. For $\Delta\Omega=\cos b\ \Delta b\ \Delta l$, where $\Delta b=\frac{\pi}{720}$ and $\Delta l=\frac{2\pi}{1440}$, and $d=d_{max}$ from the previous exercise, the integral in (8) yields the number of progenitor stars in the surveyed volume, towards a given direction. 

Well tested models for the stellar luminosity function and the stellar number density are provided in a study done by \citet{bs1980} (B\&S, hereafter). Proposed by the authors, their model compares satisfactorily to observations made in seventeen different regions on the sky. Thus, this study adopts the B\&S model with all the underlying assumptions. The B\&S stellar luminosity function is based on starcounts from a subsample of stars with known distances and apparent magnitudes. It is derived as a fit to observations in the well studied apparent magnitude range of $+4\leq m_V \leq +22$. For $-6\leq M_V \leq 15$
\begin{equation}
\Phi_M(M_V) = \frac{10^{\beta(M_V-M_V')}}{\left[1+10^{-(\alpha-\beta)\delta(M_V-M_V')}\right]^{\frac{1}{\delta}}}
\end{equation}
\[
M_V'=+1.28,\ \alpha=0.74,\ \beta=0.04,\ \frac{1}{\delta}=3.40
\]

For the number density, the B\&S model assumes an exponential distribution of stars with all the quantities derived from observations in the solar neighborhood. Since stars are born out of hydrogen gas, the distribution of massive stars is found to very well trace HI in the galaxy, but with different scale heights.
\begin{equation}
n_{\star}(r, l, b) = n_{\star\circ}\exp\left[-\frac{R(r, l, b) - R_{\circ}}{h_R}-\frac{Z_{\odot} + r\sin |b|}{h_z}\right]
\end{equation}
In the above, $n_{\star\circ} = 4.03\times10^{-3}\mbox{ pc}^{-3}$ represents the local stellar number density. B\&S propose a radial scale length for the disk $h_R\sim3.5$ kpc based on a typical HI model, however, this study adopts the same scale length prescribed for HI distribution with $h_R=0.35R_{\circ}=2.8$ kpc. For the vertical scale height, B\&S rely on the estimates given by \citet{schmidt1959} which are compounded of many investigations, undertaken by various authors (see references therein), performed roughly by the middle of the century. Considering the availability of new and better data in the past couple of decades, it is wise to seek new estimates of the scale heights. Such a study was carried out by \citet{reed2000} where he finds a vertical scale height of $h_z=45$ pc for massive OB stars in the solar neighborhood.
\subsubsection*{The Gould Belt Enrichment}
A local distribution of massive stars forms a ring like structure called the Gould Belt, after B. A. Gould who studied its orientation to the galactic disk in 1874. As mentioned previously, the seven XDINS which are relatively close by $\la 400$ pc are associated with the Gould Belt. \citet{grenier2004}, in her study of the Gould Belt structure, reports that the Sun happens to be crossing the belt which is inclined to the galactic midplane at an angle of $i=17^{\circ}$ with its center 104 pc away from the Sun towards the galactic longitude of $l=180^{\circ}$. Furthermore, the disk extends to a distance from its center of $\sim 500$ pc and appears to be devoid of any massive stars within the inner $\sim 150$ pc. 

Since the number density of massive stars belonging to the Gould belt is roughly twice that of the disk, this enriched component must be taken into account while conducting starcounts \citep{torra2000}. Also, it should be noted that the Gould belt enrichment is important only for the nearby XDINS and does not significantly affect the starcounts in the case of the AXPs. This study assumes a constant surface density model for the Gould belt with a Gaussian form for the distribution above and below the disk.
\begin{equation}
\Sigma_{\circ} = \int\rho_{\circ}\exp\left(-\frac{z'^2}{2H_{z'}^2}\right)dz',\hspace{1em}150\mbox{ pc}\la d_G \la 500\mbox{ pc} 
\end{equation}
Integrating (11) from $-\infty$ to $\infty$ yields a relation for the scale height
\begin{equation}
H_{z'} = \frac{\Sigma_{\circ}}{\sqrt{2\pi}\rho_{\circ}}
\end{equation}
With the assumption that most stars lie in the disk rather than away from it, the surface density can be expressed such that $\Sigma_{\circ} = \rho_{\circ}h$ where $h$ is the thickness of the disk. There isn't much consensus on what the thickness of the Gould Belt is, we adopt a value of $\sim 60$ pc \citep{grenier2004}. Then, from (12) the scale height $H_{z'} = 24$ pc. The radial distance from the center of the Gould disk can be expressed as follows.
\begin{equation}
d_G(r,l,b) = \left[(r\cos b\sin l)^2 + \left(\frac{r\cos b\cos l + 0.104}{\cos i}\right)^2\right. 
\end{equation}
\[
\hspace{16em}+ \left.\left(\frac{r\sin b}{\cos i}\right)^2\right]^{\frac{1}{2}}
\]
Following the observation by \citet{torra2000} that $\rho_{\circ} \approx 2n_{\star}$, the Gould belt can be incorporated into (10) as such,
\begin{equation}
n_{\star}(r, l, b) = n_{\star\circ}\left[1 + 2\exp\left\{-\frac{1}{2}\left(\frac{r\sin b}{H_{z'}\cos i}\right)^2\right\}\right]
\end{equation}
\[
\hspace{3em}\times\exp\left[-\frac{R(r, l, b) - R_{\circ}}{h_R}-\frac{Z_{\odot} + r\sin |b|}{h_z}\right]
\]
for $150\mbox{ pc }\la d_G \la 500\mbox{ pc}$.
\subsection{Birthrate}
To deduce the total number of AXPs or XDINSs in the Galaxy, the number of sample sources must be scaled accordingly. This scaling factor is the ratio of the total number of massive stars in the Galaxy to that found in the surveyed volume. The former is calculated for a heliocentric distance of $r = 25$ kpc to reach the very edge of the Milky Way. Then, the number of massive stars found for each AXP/XDINS are averaged and then scaled.
\begin{equation}
N_{AXP} = N_{total}\sum_{i=1}^{k}\frac{1}{N_i}
\end{equation}
where $k$ is the observed population of AXPs. $N_i$ and $N_{total}$ represent the number of massive stars in the probed volume for each source and the total number of massive stars in the whole Galaxy, respectively. Assuming that the populations of AXPs and XDINSs are in a steady state, then their birthrate is calculated using their spindown ages $\tau_c$. 
\begin{eqnarray}
\tau_c &=& \frac{P}{2\dot{P}}\\
\beta &=& N_{total}\sum_{i=1}^{k}\frac{1}{N_i\tau_{c_i}}
\end{eqnarray}
\section{Results and Discussion}
Maximum distance plots for AXP 4U 0142+61 and XDINS RBS 1223 are provided below. Fig 3 illustrates the maximum distances to which an object with the spectral characteristics similar to AXP 0142+61 and XDINS RBS 1223 could be detected for criterion B. Strikingly, the curve represents the disk of the Galaxy where the gas and dust limits the ability of the telescope to peer deep into the midplane. It is not surprising to note that the radiation from bright X-ray sources is primarily absorbed in the plane of the disk, whereas the light travels unimpeded immediately above and below it. 
\begin{figure}
\begin{center}
\includegraphics[scale=0.55]{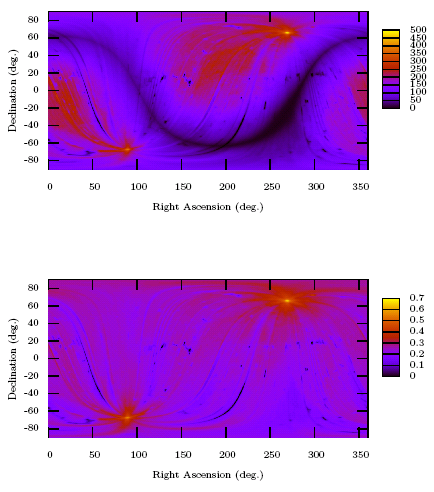}
\caption{Maximum distance plots for (top) AXP 4U 0142+61 and (bottom) XDINS RBS 1223. The shading represents the distance in units of kpc. Noticeable in the top figure are the disk and the central bulge of the Galaxy. This structure is missing in the bottom figure due to the low luminosity of XDINSs that makes them undetectable even before probing deeply into the galactic midplane. The upper limit on the distance in the bottom figure lends further support to this fact. In addition, the figures illustrate the sensitivity variations caused by inconsistencies in the exposure times as described earlier, and the bright spots mark the ecliptic poles.}
\end{center}
\end{figure}
The knowledge of the maximum surveyed distance at each point on the
sky enabled us to calculate the number of massive, OB stars in the
probed volume. For example, starcounts for AXP 4U 0142+61 yielded
267,575 and 229,967 progenitor stars for limiting count rate A and B,
respectively. Similarly, we found 175 progenitor stars for both A and
B limiting count rates in the case of the XDINSs. Furthermore,
integrating the models to the outer regions of the Galaxy, we found
523,196 massive OB stars without the Gould belt enrichment. Upon its
addition, the total number increased to 523,694. The final results of
this exercise are provided in Table 1 with estimates of the
population, and birthrates, for both classes of NSs. Also listed in
the table are the population and birthrate statistics of SGRs
calculated using (15) and (17) with the assumption of $N_i =
N_{total}$; however, with so few objects, this value provides only an
order-of-magnitude estimate.
\begin{table}
\caption{Listed here for criterion (A) and (B) are the total number of
  objects in the Galaxy, and the corresponding birthrate for both AXPs
  and XDINS. Also, the SGR statistics are provided for comparison 
\citep{kouveliotou1998}.}
\begin{tabular*}{1.0\columnwidth}{@{\extracolsep{\fill}}lrr}
\hline
&A&B\\
\hline
\hline
$\beta_{AXPs}$ (per century)&0.20&0.22\\
$N_{AXPs}$&12&14\\
$\beta_{XDINSs}$ (per century)&2.1&2.1\\
$N_{XDINSs}$&22,932&22,932\\
\hline
$\beta_{SGRs}$ (per century)&$\sim 0.1$\\
$N_{SGRs}$&3\\
\hline
\end{tabular*}
\end{table}

Previously, many workers have tried to answer the question asked in this study, and a possible lower limit of 0.1 magnetars per century has been suggested \citep{kouveliotou1998,paradijs1995}. This study finds the birthrate of magnetars to be 0.22 per century, an estimate that is in good agreement with the proposed lower limit. Essentially, this is the birthrate of active magnetars, in this case of active AXPs. If magnetars spend a large fraction of the time in quiescence and, consequently, evaded detection in the RASS, then the birthrate will be correspondingly larger. However, since the birthrate is only a factor of eight smaller than the galactic supernova rate \citep{reed2005}, the duty fraction of these objects cannot be too small. In addition, from table 1 one will immediately be struck by the relatively small difference in birthrates between the AXPs and the SGRs. From this we infer like others \citep{zhang2002,wt2006} that it is possible that the AXPs represent a later evolutionary stage of the SGRs. Furthermore, the birthrate found in this study cannot be assumed as the upper limit with utmost certainty due to some of the factors described below.

Firstly, it can be argued that the model assumed for the distribution of HI in the galaxy may be too simplistic. It is very well known that our galaxy, being spiral in its morphology, has substantial density enhancements along the spiral arms. The existence of such a structure can't be ignored, yet the model, for this study, has been kept deviod of it. Addition of the spiral structure in the HI model would act to increase the average sum $\frac{1}{k}\sum_{i=1}^k\frac{1}{N_i}$ as the volume will now be less than previously calculated, effectively, reducing the number of massive stars in the probed volume. Consequently, the birthrate estimate will be augmented. Nevertheless, it should be noted that the addition of the spiral arms must also increase the number of massive stars as these density enhancements are their likely birthplace. After weighing both arguments, one can probably conclude that the result won't be affected considerably. As an example, a simple error analysis for the AXP birthrate, based on the counting error of the sample $1/\sqrt{5}\sim 40 \%$ and a naive estimate of the error in the spindown ages $\sim 50 \%$, amounts to an uncertainty of $\sim 0.2$ per century. Similarly, we estimate an uncertainty of $\sim 1$ per century for the XDINSs' birthrate.   

Secondly, the form adopted for the distribution of massive stars in the Galaxy is similar to that of HI distribution with a radial scale height of 0.35$R_{\circ}$. However, the vertical scale height was adopted from Reed's \citeyearpar{reed2000} model. In his estimate of the number of massive O3-B2 stars inside the solar radius, Reed gives a number of $\sim$ 200,000 stars \citep{reed2005}. Scaling this number out to the radius of the Galaxy with $R_G \sim$ 15 pc by $\left(\frac{R_G}{R_{\circ}}\right)^2$, since most of these stars belong to a thin disk, yields a result of $\sim$ 700,000 massive stars. Comparing this number to what was obtained for the simple model assumed for this study, a conservative estimate of 523,196 falls short of Reed's estimate. This discrepancy may cause a further reduction in the calculated birthrate.

Thirdly, \citet{colpi2000} argue that the characteristic spindown ages of magnetars inferred from (16) may be larger than their true ages. The basis of their argument rests on the hypothesis that the magnetic fields, usually assumed to be constant, are decaying over time. If it is so, then a smaller characteristic age would give a larger birthrate. Analysis conducted upon fixing for this proposed discrepancy yields an AXP birthrate of $\sim$ 2 per century. Until the issue of the underlying mechanism for magnetar production is resolved, it wouldn't be unwise to assume the characteristic spindown ages as the true ages of the pulsars.

This study finds the birthrate of XDINSs to be $\sim$ 2 per century. Unfortunately, this result was obtained while generalizing the characteristic age ($\sim$ 1.7 Myr) of just two sources (RX J0720.4-3125 and RBS 1223; \citealp{kaplan2005a,kaplan2005b}) to that of the whole sample; furthermore \citet{heyl1998} find cooling ages for these sources of about 0.5 Myr. For want of a better determination we have taken the mean of these values $\sim$ 1.1 Myr to be the typical age of these sources. Under these assumptions, the estimated birthrate is in excellent agreement to the value proposed by \citet {popovetal2006} for both the XDINSs and the RRATSs.

After having obtained the population size estimates for both the AXPs and the XDINSs, it is now possible to find any evolutionary affiliation among the different classes of NSs, simply by comparison of these statistics. We find that there are as many as 2.8 times more AXPs than already observed in our galaxy. This corresponds to a population of 14 AXPs. Including the three SGRs, assuming that all of them are observable irrespective of their location, the total magnetar population amounts to 17 sources in the Galaxy. Similarly, we find that there are $\sim$ 22,000 XDINSs in our galaxy. Although the birthrate of XDINSs roughly matches the upper limit of that of the normal radio pulsars, the number of normal radio pulsars (70,000 - 120,000; \citealp{vranesevic2004}) is larger than that of the XDINSs due to their larger typical ages. We find that the birthrate of the AXPs and SGRs falls short of the RRATSs (and XDINSs) so that they cannot account for the RRAT population unless there is a large population of transient AXPs, a similar conclusion to that found by \citet{popovetal2006} in their analysis of magnetar, XDINS and RRAT birthrates. 

On the other hand, the magnetar birthrate is in good accord with the
HBRP birthrate of $\sim$ 0.2 per century, however, the magnetar
population again falls short of the HBRP population of $\sim 187\pm
103$ \citep{vranesevic2007}.  This difference in population simply
reuslts from the typically larger ages of the HBRPs relative to the
AXPs. Considering other instances, such as the X-ray emission from
many HBRPs, the discovery of radio emission from the transient AXP XTE
J1810-197 \citep{camilo2006}, and the comparability of spin periods
between the two classes, place the HBRPs in a circle of suspicion for
having some connection with the magnetars.
\section{Recommendations}
Recently, hard X-ray emission has been detected from three AXPs (4U 0142+61, 1E 1841-045, 1RXS J170849.0-400910) by INTEGRAL space telescope operating in the 15 keV - 10 MeV energy range \citep{gotz2006}. Furthermore, an all sky survey, conducted by using the coded mask IBIS telescope on board INTEGRAL, has identified various high energy galactic sources, a few among them ($\sim$ 42) are yet to be identified \citep{krivonos2007}. Considering the results of this study and noting the possibility of detecting hard X-ray emission from the AXPs, as well as being highly optimistic, we would like to extend our speculation that some of the unidentified sources may be those AXPs that have evaded detection in the low energy regime.
\section{Conclusion}
According to this study, up to 22\% of all supernovae explosions give birth to a magnetar. The small size of the magnetar population reflects their short lifespans during which these objects manifest themselves as being highly energetic sources in the Galaxy. The birthrates obtained while assuming decaying magnetic fields, as advocated by \citet{colpi2000}, exceed estimates of the supernova rate in the Galaxy \citep{reed2005}, thus, this study resorts to constant fields. Further timing analysis is required to gain important insight into the characteristics of XDINSs as this has been successfully achieved for only two among the seven.
\section*{Acknowledgments}
The Natural Sciences and Engineering
Research Council of Canada, Canadian Foundation for Innovation and the
British Columbia Knowledge Development Fund supported this work.
Correspondence and requests for materials should be addressed to
J.S.H. (heyl@phas.ubc.ca).  This research has made use of NASA's
Astrophysics Data System Bibliographic Services
\bibliographystyle{mn2e}
\bibliography{refs}
\label{lastpage}
\end{document}